\documentstyle[preprint,aps]{revtex}
\begin {document}
\draft
\preprint{UCI-TR 95-9}
\title{Tetraquarks: Mesons with two $b$ quarks}
\author{Myron Bander\footnote{Electronic address: mbander@funth.ps.uci.edu}
%\addtocounter{footnote}{3}%
and
A.\ Subbaraman\footnote{Electronic address: anand@funth.ps.uci.edu}}
\address{
Department of Physics, University of California, Irvine, California
92717}

\date{March\ \ \ 1995}
\maketitle
\begin{abstract}
We look for the existence of stable (by strong interactions)
tetraquarks, mesons containing two heavy antiquarks and two light
quarks, specifically a $|{\bar b}{\bar b}qq\rangle$ state. This state
is viewed as two $|{\bar b}q\rangle $ particles bound by a potential
due to the exchange of the four light pseudoscalar and vector mesons,
namely the $\pi,\ \eta,\ \rho$ and $\omega$. Couplings of these
particles are obtained by the use of heavy quark symmetry. We find that
the $I=0,\ J=1$ state is bound for a large range of parameters, while
all other states are not bound (the $I=1,\ J=0$ system may be bound for
some parameters at the extreme range of their allowed values). We
discuss the reasons why the techniques used cannot be extended to
mesons containing two $c$ quarks or a $c$ and $b$ quark.
\end{abstract}


%\narrowtext
\section{Introduction}\label{INTRODUCTION}
In addition to the usual three quark baryons and quark-antiquark mesons,
hadrons containing one or more heavy quarks may occur in other configurations,
stable against decay by strong interactions. Baryons  with one heavy
antiquark and four light quarks, {\it i.e.} pentaquarks \cite{pentaquark},
as well as baryons with two heavy antiquarks and five light
quarks, {\it i.e.} heptaquarks
\cite{heptaquark}, have been considered. Mesons containing two heavy
antiquarks and two light quarks, tetraquarks
\cite{tetraquark,Tornqvist,ManWise,Richard}, have also been studied. The
approach used in the last two references was to consider a tetraquark as two
mesons, each containing a heavy antiquark, bound to each other through the
exchange a $\pi$ meson.

In this work we shall look for bound states of mesons containing two
${\bar b}$ quarks and two light $u$ or $d$ quarks, namely a $({\bar b}{\bar
b}qq)$ system. As mentioned above we shall view such a possible system as a
bound state of two $B$ mesons, two $B^*$ mesons, or a $BB^*$ combination.
It turns out that the bound states are linear combinations of these
configurations. The binding potential will be due to the exchange, in
addition to the $\pi$, of the $\eta$,\   $\rho$ and $\omega$ mesons. The
relevant coupling constants are obtained in
Section \ref{intpot} using heavy quark symmetry \cite{HQL}. The
Schr\"{o}dinger equations and their solutions are discussed in Section
\ref{Schr}. We find that bound states may exists for the light quarks in a
symmetric color combination (presented in Section \ref{config}); this
is the opposite of the conclusions from bag model arguments
\cite{tetraquark,Richard}. For a wide range of parameters we find that a bound
state exists for the system with total isospin $I=0$ and total angular
momentum $J=1$. The range of binding energies below a possible decay
to $B$ and $B^*$ mesons is about 650 MeV to 250 MeV. The $I=1$ and
$J=0$ is unlikely to be bound.

Binding of states containing two $c$ quarks or a $c$ and a $b$ quark
are discussed in Section \ref{DD}. Our methods cannot be extended to
the study of these states. As $M_{D^*}-M_D\sim m_\pi$ and for some
charge combinations $M_{D^*}-M_D > m_\pi$ we cannot use the
multi-channel approach in which we neglect the mass difference of the
heavy pseudoscalar and vector mesons.  Diagrams where the $\pi$ is
exchanged between a $D$ and a $D^*$ result either in a weak very long
range potential or in an imaginary one, indicating that one should
treat the $DD^*$ state as a three body $DD\pi$ system.

\section{INTERACTION POTENTIAL}\label{intpot}
We shall be interested in obtaining the potential between two heavy mesons. In
the heavy quark limit \cite{HQL} the pseudoscalar and vector mesons are
degenerate in mass and the bound state problem has to be treated as a
multichannel one. As in this limit the masses of the incident and final pairs
are equal and the energy of each of these states is just the sum of the two
heavy masses, we are also allowed to use non-relativistic methods. These
approximations are adequate for mesons containing $\bar b$ quarks but, as we
will discuss, are likely to break down for systems with lighter quarks.

\subsection{Heavy-light coupling constants}\label{coupling constants}
As in Ref. \cite{ManWise} we shall use the formalism of heavy quark
symmetry and of the nonlinear chiral model\cite{HQCS} to obtain the coupling
constants for the exchange of light mesons.  With $B^a$ and $B^{*a}_\mu$
denoting the fields for the pseudoscalar $B$ and vector $B^*$, $a$
being an isotopic spin index, we can define a heavy,
fixed four--velocity $V^\mu$ field $H^a$,
\begin{equation}
H^a = \left( {{1 + V\cdot \gamma}\over 2} \right) ( i \gamma^5 B^a
    + \gamma \cdot B^{*a}_\mu), \, \,\,  {\bar H} = \gamma^0 H^\dagger
\gamma^0 \, .
\end{equation}
The free Lagrangian for $H$ is
\begin{equation}
{\cal L}_{free}= -iM\,V^\mu \mbox{\rm Tr}\left( H\partial_\mu {\bar
H}\right)\, .\label{Lfree}
\end{equation}
The light Goldstone pions and the $\eta$ are described by a $U(2)$
matrix
\begin{equation}
\Sigma=\exp \left \{ {{{\sqrt 2}i}\over {F_\pi}}
(\tau\cdot\pi+\eta) \right\} \, ,
\end{equation}
with $F_\pi = 132$ MeV. Chirally invariant Lagrangians can be written
with the help
of the square root of the $\Sigma$ field, $\xi=\Sigma^{1/2}$. We define two
combinations
\begin{equation}p_\mu = {i\over 2}(\xi \partial_\mu \xi^\dagger -
\xi^\dagger \partial_\mu
\xi), \qquad v_\mu =  {i\over 2}(\xi \partial_\mu \xi^\dagger + \xi^\dagger
 \partial_\mu \xi)\, \, .\label{defap}
\end{equation}
The interaction of heavy mesons with the light ones is given by
\begin{equation}
{\cal L}_d = M \, d \, \mbox{\rm Tr} \left(H\gamma^\mu \gamma^5 p_\mu {\bar
H}\right)\, . \label{picoupling}
\end{equation}
The coupling constant $d$ is not very well known; from a single pole fit to
the data on $D \rightarrow K$ and $D \rightarrow K^*$ semileptonic decays
\cite{dvalue}, one obtains $d \approx 0.53$. A straight forward use of the
static $SU(6)$, or more precisely $SU(4)$ symmetry gives $d=1$; reducing
this by the ratio $(g_A/g_V)/(5/6)$ results in $d=0.7$. The value
$d=0.53$, which we shall use, is a conservative estimate. If we include
the strange quark content of the $\eta$ and $\eta-\eta '$ mixing the
heavy meson-$\eta$ coupling has to be reduced to $d'$; using the known
$\eta-\eta '$ mixing angle of $11^\circ$ we find that $d'/d\sim 0.4$.

The light vector meson fields, $\rho_\mu$ and $\omega_\mu$, are
combined as
\begin{equation}
\rho_\mu = {1\over {{\sqrt 2}}}(\rho_\mu\cdot \tau + \omega )\, ,
\end{equation}
and we also define a field strength tensor
\begin{equation}
F_{\mu \nu}(\rho) = \partial_\mu \rho_\nu - \partial_\nu \rho_\mu
- i {\tilde g}[\rho_\mu, \rho_\nu] \, .
\end{equation}
The width of the $\rho$ sets ${\tilde g}=3.93$.
The coupling of these vector mesons is accomplished by modifying
\cite{dvalue} ${\cal L}_{free}$, Eq. (\ref{Lfree}) to
\begin{equation}
{\cal L}_{kinetic}= -iM\,V_\mu Tr\left[ H\{\partial_\mu - i\alpha {\tilde g}
\rho_\mu - i(1-\alpha)v_\mu\} {\bar H}\right]\, ,
\end{equation}
where $v_\mu$ is defined in Eq. (\ref{defap}). This term is invariant
under chiral transformations. Vector dominance suggests that
$\alpha \approx 1$, which is the value we use in this work.  In
addition we add a magnetic moment term
\begin{equation}
{\cal L}_c = {{-icM}\over {m_v}} Tr \left[ H \gamma^\mu \gamma^\nu
F_{\mu \nu}(\rho) {\bar H}\right] \, .
\end{equation}
$m_v$ is a fiducial light vector meson mass, which appears from
dimensional considerations.
The coupling constant $c$ is constrained to \cite{Jain}
\begin{equation}
2.27 < c/d < 3.75  \, . \label{crange}
\end{equation}

Phenomenology of the nucleon-nucleon potential \cite{nuclnucl}
requires the exchange of a $0^+(0^{++})$ particle with a mass of $\sim 550$
MeV. This represent the effective potential due to the exchange of two
pions in box diagrams with isobars in the direct channel and the
exchange of the broad $\pi-\pi$ resonance $f_0(980)$. In our case, by
considering both the scalar and vector mesons in a multichannel
formalism, the box diagrams are automatically included and the exchange of the
$f_0(980)$ contributes to that part of the potential whose range is
shorter than that which we consider explicitly; hence we will not include
any $0^+(0^{++})$ exchanges.

\subsection{Interaction potential}
Using the coupling constants determined in the previous subsection it
is straightforward to find the interaction potential. Let $\mbox{{\bf S}}^i$
denote the spin of the light quark in the heavy meson, $i=1,2$,
while {\bf S} and
{\bf I} denote the total spin and isospin of the two light quarks. The
position space potential is a combination of central, spin-spin and tensor
terms,
\begin{equation}\label{potential}
V({\mbox{\bf x}})= V_0({\mbox{\bf x}}) + ({\mbox{\bf S}}^1 \cdot {\mbox{\bf
S}}^2)V_S({\mbox{\bf x}}) +
\left( ({\mbox{\bf S}}^1 \cdot {\hat {\mbox {\bf x}}})\,
({\mbox{\bf S}}^2 \cdot
{\hat {\mbox {\bf x}}})-  {1\over 3}{\mbox{\bf S}}^1 \cdot {\mbox{\bf S}}^2
\right) V_T({\mbox{\bf x}}) \, ,\label{pot1}
\end{equation}
where
\begin{eqnarray}
V_0({\mbox{\bf x}}) &= & {{{\tilde g}^2}\over {4\pi}} \left[
({\mbox{\bf I}}^2 -
{3\over 2}) {{e^{-m_\rho r}}\over r} + {1\over 2}{{e^{-m_\omega r}}\over r}
\right]\, ,\nonumber\\
V_S({\mbox{\bf x}})& =& {{d^2}\over {\pi F_\pi^2}}\left[(\mbox{{\bf I}}^2 -
{3\over 2}) Y(r,m_\pi) + {1\over 2}\frac{d'^2}{d^2}Y(r,m_\eta)\right]
+ {{4c^2}\over {\pi m_{v}^2}}\left[(\mbox{{\bf I}}^2 - {3\over 2})
2Y(r,m_\rho) + Y(r,m_\omega)\right] \, ,\nonumber\\
V_T({\mbox{\bf x}}) &= &{{d^2}\over {\pi F_\pi^2}}\left[(\mbox{{\bf I}}^2 -
{3\over 2}) F(r,m_\pi) + {1\over 2}\frac{d'^2}{d^2}F(r,m_\eta)\right]
- {{4c^2}\over {\pi
m_{v}^2}}\left[(\mbox{{\bf I}}^2 - {3\over 2}) F(r,m_\rho) + {1\over 2}
F(r,m_\omega)\right]\, , \nonumber\\  \label{pot2}
\end{eqnarray}
and
\begin{eqnarray}
F(r,m) &=& e^{-mr}\left( {3\over{r^3}}+{{3m}\over{r^2}}+{{m^2}\over r}\right)
\, , \nonumber\\
Y(r,m)&=&{{m^2}\over {3r}}e^{-mr}\, .
\end{eqnarray}
The $1/r^3$ part of the tensor force, $V_T$, is too singular to permit a
solution of the resulting Schr\"{o}dinger equation; in addition there is a
delta function at the origin lurking in that potential. As we have not taken
into account the exchange of mesons whose mass is greater than that of the
light vector mesons we will, as in Ref. \cite{ManWise}, flatten the
potentials at some $r_c$; we will choose $r_c= 1/m_\rho$; we will discuss
how the results depend on $r_c$ for $r_c>1/m_{\rho}$.

\section{BOUND STATE CONFIGURATIONS}\label{config}
In order to obtain a bound state we require that the long range part of the
potential ($\pi$ exchange) be attractive. From Eqs. (\ref{pot1},\ref{pot2})
we see that the light quark configurations must be either $\mbox{\bf I}=1$,
$\mbox{\bf S}=0$ or $\mbox{\bf I}=0$, $\mbox{\bf S}=1$; as expected the
greatest attraction is in an orbital $S$-wave state. As the
space-spin-isospin of the light quarks is antisymmetric they must be in a
symmetric color state, the {\bf 6}; this implies that the two ${\bar
b}$ quarks are in the {\bf 6}$^*$, forcing them to be in a spin
singlet state. Thus the possible bound states are either a $I=1,\,
J=0$ state or a $I=0,\, J=1$ state. These are linear combinations of $B$'s and
$B^*$'s:
\begin{eqnarray}
|I=1,I_3; J=0\rangle &=&\frac{1}{2}\langle 1,I_3|1,m;1,m'\rangle
\nonumber\\&{}&\left (
|B^{*m}_{+1}B^{*m'}_{-1}\rangle +|B^{*m}_{-1}B^{*m'}_{+1}\rangle -
|B^{*m}_0B^{*m'}_0\rangle +|B^mB^{m'}\rangle \right ) \, ,\nonumber\\
|I=0; J=1, J_3=1\rangle &=&\frac{1}{2}\langle0,0|1,m;1,m'\rangle
\nonumber\\ &{}&\left (
|B^{*m}_{+1}B^{*m'}_{0}\rangle +|B^{*m}_{0}B^{*m'}_{+1}\rangle -
|B^{*m}_{+1}B^{m'}\rangle +|B^mB^{*m'}_{+1}\rangle \right )\, ,
\label{decomp}
\end{eqnarray}
with similar expressions for the other two $J=1$ states.

The expectation value of the mass operator in these states is twice the spin
averaged mass
\begin{equation}
\langle {\cal M}\rangle=2M_{\mbox{\rm eff}}=2\,\frac{3M_{B^*}+M_{B}}{4}\,
.\label{effmass}
\end{equation}
Eqs. (\ref{decomp}) tell us into what combinations of $B$'s and $B^*$'s these
states may decay.

\section{SCHR\"{O}DINGER EQUATIONS AND SOLUTIONS}\label{Schr}
\subsection{Eigenvalue problem}
To order $1/M$ the Hamiltonian of this system is
\begin{equation}
H=2M_{\mbox{\rm eff}}-\frac{1}{M_{\mbox{\rm eff}}}\mbox{\bf $\nabla$}^2
+V({\mbox{\bf x}})\, .\label{Hamilt}
\end{equation}
A basis for the system is provided  by the states
$|II_3\rangle \,|SS_3\rangle \,|lm\rangle $, where
$l,m$ are the orbital angular momentum quantum numbers of the bound
system. Note that $l,m$ includes the orbital angular momentum of
the heavy quarks as well.
Since the potential is rotationally invariant, the total angular momentum
\begin{equation}
{\mbox{\bf J}}= {\mbox{\bf S}}+ {\mbox{\bf L}}
\end{equation}
is a good quantum number and can be used to label the eigenstates.
The $J=0$ (singlet) state is written as
\begin{equation}
\Psi_{J=0} =
{{\psi(r)}\over r} \,Y_{00}\,
|s=0 \rangle \, .
\end{equation}
The $J=1$ (triplet) state is a linear combination of
$|S=1,\,l=0 \rangle$ and
$|S=1,\,l=2 \rangle $ and can be written as
\begin{equation}
\Psi^{J_3}_{J=1} = {{u(r)}\over r}\,{\cal Y}^{J_3}_{1;0,1} +
{{w(r)}\over r}\,{\cal Y}^{J_3}_{1;2,1}
\, ,
\end{equation}
where ${\cal Y}^{J_3}_{J;l,S}$ is a generalized spherical harmonic
obtained by combining spin and orbital angular momentum.
Both the singlet and triplet states must further be supplemented by
an isospin label, with $I=0$ or 1.

With $\epsilon=E-2M_{\mbox{\rm eff}}$, the eigenvalue equation for the
singlet case is
\begin{equation}
\psi'' = M_{\mbox{\rm eff}}\left(V_0 -{3\over 4}V_S -\epsilon \right)\,\psi
\, ;\label{singlet}
\end{equation}
prime denotes differentiation with respect to $r$. For the spin triplet case
the tensor force couples $l=0$ to $l=2$ and we get two coupled differential
equations,
\begin{eqnarray}
u''&=& M_{\mbox{\rm eff}}\left[ \left(V_0 + {1\over 4}\,V_S
-\epsilon\right)\,u + {1\over {3{\sqrt 2}}}
\,V_T \,w \right] \, ,\nonumber \\
w'' &=& M_{\mbox{\rm eff}}\left[ \left( V_0 + {1\over 4}\,V_S + {6\over
{M_{\mbox{\rm eff}}\,r^2}} -{1\over 6}\,V_T -\epsilon\right)\,w+
{1\over {3{\sqrt 2}}}\,V_T\,u \right] \, . \label{triplet}
\end{eqnarray}

\subsection{Numerical results}

We now consider the numerical solutions to Eqs. (\ref{singlet}) and
(\ref{triplet}) for the energy eigenvalue $\epsilon$; $d$ is fixed at
0.53 and the $B,\, B^*$ masses are taken to be 5.279 GeV and 5.325 GeV
respectively. $c/d$ is varied between 2.27 and 3.7.

For the singlet case, we find that there are bound
states only in the $I=1$ channel, forcing the color wavefunction to
be in the symmetric {\bf 6} representation, as mentioned earlier.
We observe from the decomposition Eq. (\ref{decomp})
of the singlet wavefunction
that the lightest constituents the state can decay into is two $B$ mesons.
Hence the true binding energy is
\begin{eqnarray}
\epsilon_B &&= 2 M_{\mbox{\rm eff}} - 2M - |\epsilon| \nonumber \\
&& = 69 {\mbox{ MeV}} - |\epsilon| \, ;
\end{eqnarray}
the state is bound if $\epsilon_B$ is negative. Binding occurs only
for $c/d\ge 3.6$ and $\epsilon_B=40$ GeV at the extreme of the
allowed range, $c/d=3.7$. Hence we conclude that the singlet
configuration is unlikely to be bound.

We now move to the triplet configuration. Here, the lightest constituents
the state can decay into is a $B$ and a $B^*$ meson. hence the true
binding energy is given by
\begin{eqnarray}
\epsilon_B &&= 2 M_{\mbox{\rm eff}} - M -M^* - |\epsilon| \nonumber \\
&& = 23 {\mbox{ MeV}} - |\epsilon| \, .
\end{eqnarray}
Fig.\ 1 summarizes the numerical results for the allowed range of $c/d$.
It is clear from the figure that there is a stable bound state over
the entire allowed range of the coupling constant $c$.

One interesting issue to explore is the sensitivity of the binding
energy to the cutoff $r_c$, below which the potential is flattened.
Fig. 2 shows the dependence of $\epsilon_B$ on $r_c$, for a central value
of the ratio $c/d = 3$. The lightest particle above the vector mesons
which can contribute to the potential
is the $f_0(980)$, or the $\eta^\prime(958)$. These correspond
to a value of $r_c$ in the neighborhood of about 1 GeV$^{-1}$. We see from
Fig.\ 2 that a cutoff less than about 1 GeV$^{-1}$ leads to a very high
binding energy; indeed, the non-relativistic formulation of the system
breaks down. Hence the potential derived here cannot be trusted for
values of the cutoff much less than 1 GeV$^{-1}$. We note that
$r_c=1/m_{\rho}$ is a conservative estimate for determining whether
binding does or does not occur.

\section{$D-D$ and $D-B$ systems}\label{DD}
We repeated the calculations for smaller reduced masses to see what
happens for the $D-D$ and $D-B$ systems. From Fig.\ 3 we could claim
that both of these will be bound for a triplet light quark spin
configuration. This interpretation is too naive. As has been
emphasized in Ref. \cite{Tornqvist} the heavy quark limit cannot be
use for the study of binding of charmed mesons. For $\pi$ exchange the
range of the potential is nor $1/m_\pi$, but $1/\mu$ with
$\mu=\sqrt{m_{\pi}^2-(M_D^*-M_D)^2}$. This value is close to zero, and
for several charge combinations negative. (For the $B-B$ system this
gives a 6\% correction to the range of the $\pi$ exchange potential.)
Thus the breaking of heavy quark symmetry and even the breaking of
isospin invariance will be crucial to the study of this system. As in
the nucleon-nucleon case we could look at the single channel $D-D$
problem.  As there is no one-$\pi$ exchange potential for this
configuration, this system is unlikely to be bound. The $D-D^*$ mass
difference would force us to treat this as a three body $D-D-\pi$ system.

\section*{ACKNOWLEDGEMENTS}
This research was supported in part by the National Science Foundation under
Grant PHY-9208386. We wish to thank Professor S. Nussinov for discussions.

\nobreak
%\newpage

\begin{figure}
\caption{Binding energy, $\epsilon_B$,  of the $I=0,\ J=1$ state as a
function of the
magnetic vector meson coupling $c/d$.}
\end{figure}
\begin{figure}
\caption{Binding energy, $\epsilon_B$, of the $I=0,\ J=1$ state as a
function of the cutoff for $c/d=3.03$.}
\end{figure}
\begin{figure}
\caption{Eigenvalue of the Schr\"{o}dinger Equation, $\epsilon$, of the
$I=0,\ J=1$ state as a function of the reduced mass for $c/d=3.03$ and
$r_c=1/m_{\rho}$.}
\end{figure}
%\narrowtext
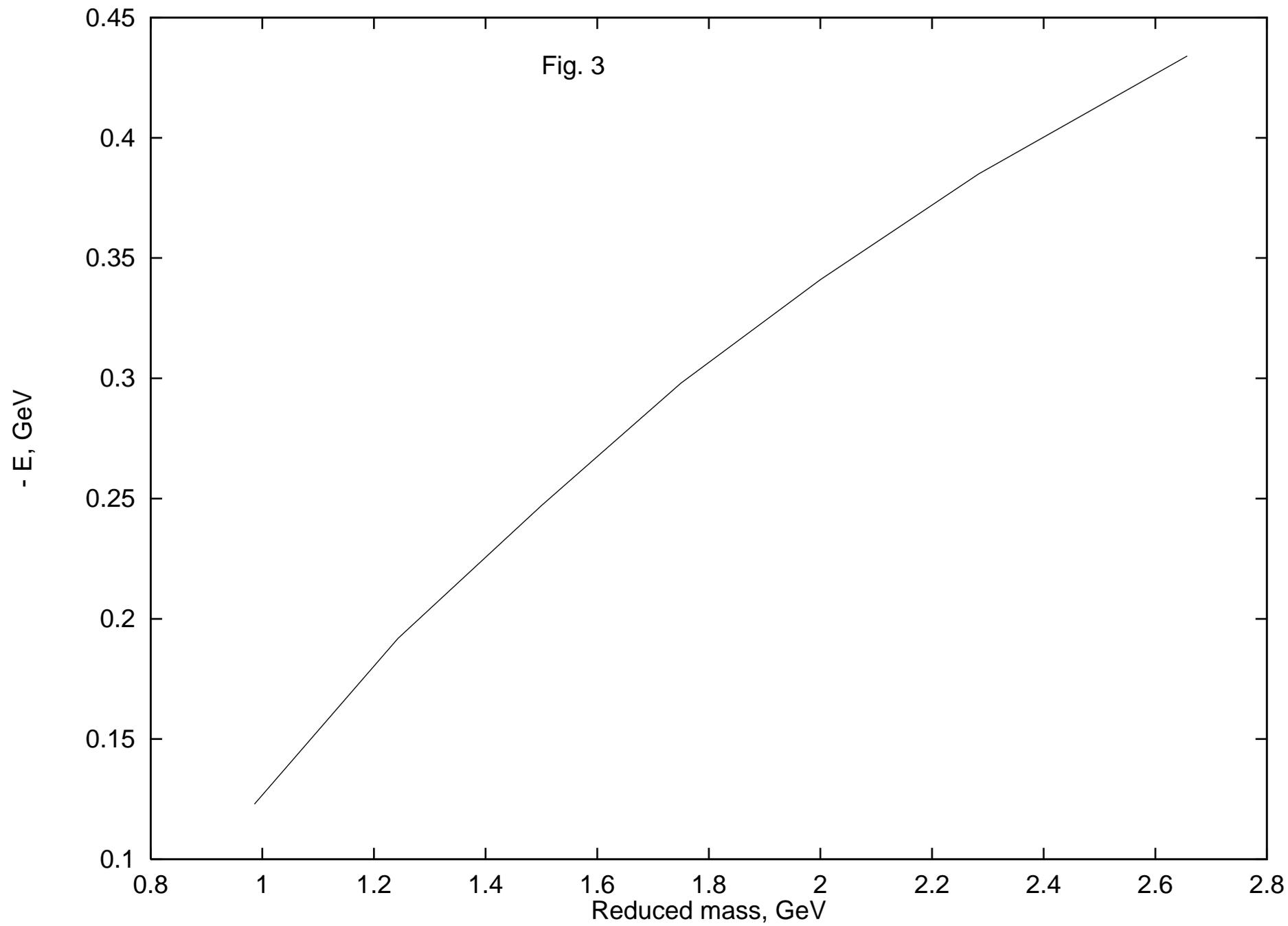
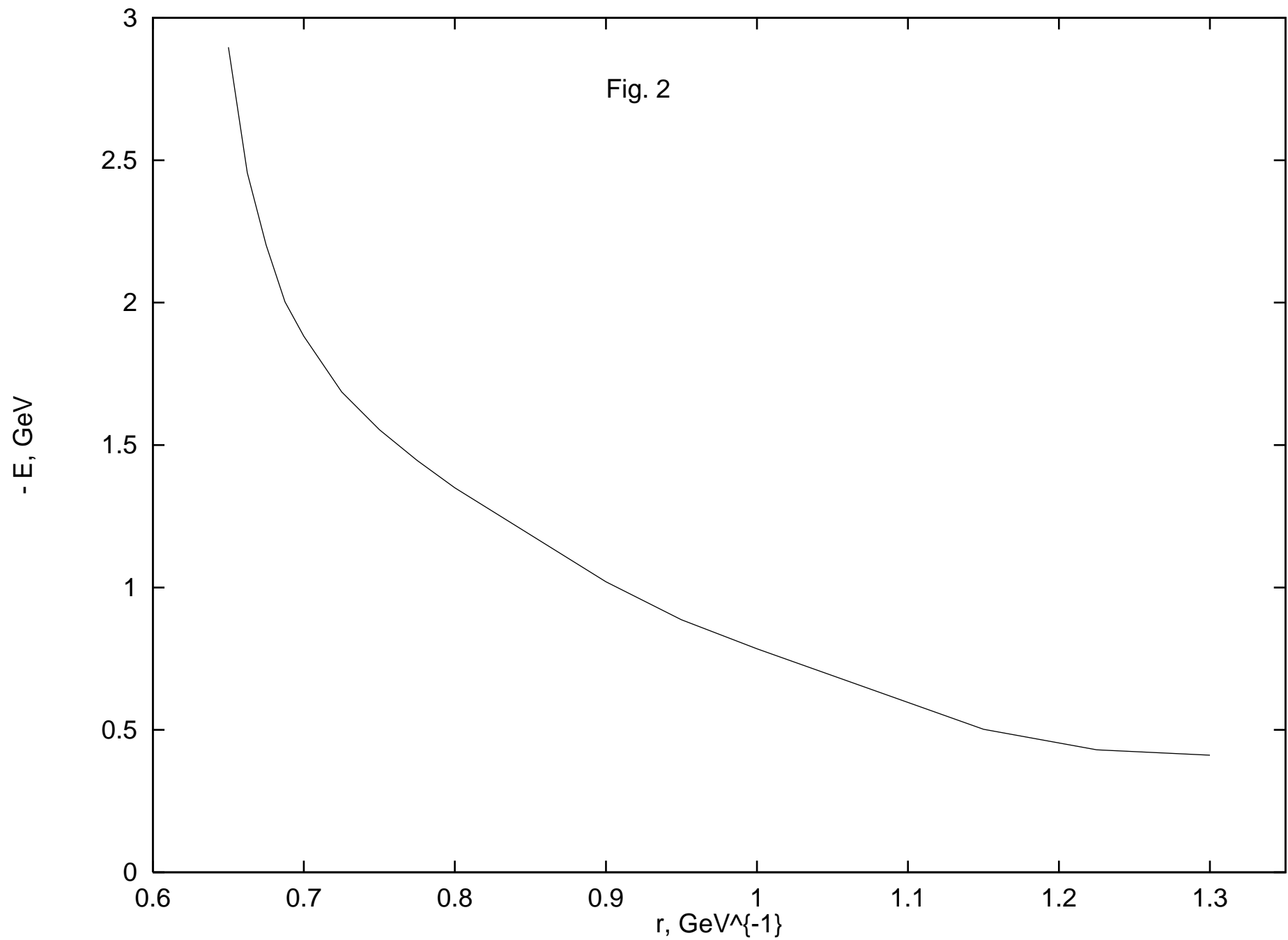
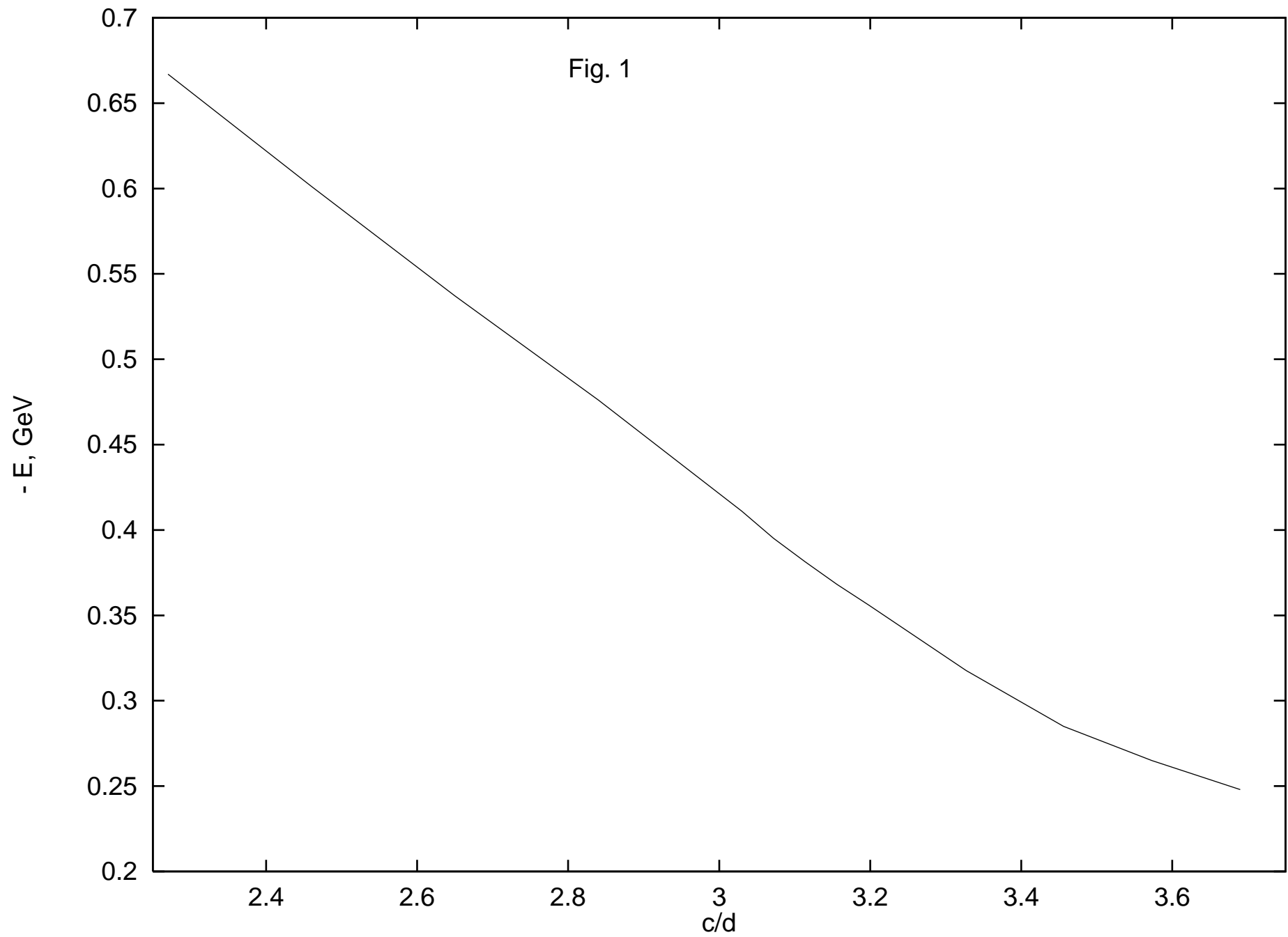
\end{document}